\documentclass[10pt]{iopart}

\usepackage{graphicx}
\begin{document}

\title[Argon force field revisited: a molecular dynamic study]
{Argon force field revisited: a molecular dynamic study}

\author{Jos\'e Guillermo M\'endez-Berm\'udez$^1$, Iv\'an Guill\'en-Escamilla$^1$, 
Gloria Arlette M\'endez-Maldonado$^2$, and Jos\'e Abundio Daniel Alva-Tamayo$^1$}

\address{$^1$Centro Universitario de los Valles, Universidad de Guadalajara, 
Carretera Guadalajara-Ameca Km. 45.5, Ameca, 46600, Jalisco, M\'exico}
\address{$^2$Centro Universitario de Ciencias Exactas e Ingenier\'ias, 
Universidad de Guadalajara, Blvd. Marcelino Garc\'ia Barrag\'an No. 1412, 
Guadalajara, 44430, Jalisco, M\'exico}
\ead{jose.bermudez@academicos.udg.mx,ivan.guillen@academicos.udg.mx}
\vspace{10pt}

\begin{abstract}
We report the improvement of five argon force fields by scaling Lennard-Jones energy 
($\epsilon$) and distance ($\sigma$) parameters to reproduce liquid-vapor 
phase diagram and surface tension simultaneously, with molecular dynamics. Original force 
fields reproduce only liquid-vapor phase diagram among other properties except surface 
tension. Results showed that all force fields converge in a nearby region in the 
$\epsilon$-$\sigma$ phase space, which is different from the original values. This study 
gives the intervals where the numerical values of $\epsilon$ and $\sigma$ reproduce 
both properties mentioned above.  

\end{abstract}


\submitto{\JPCM}
\maketitle

\begin{figure}[h!]
\centering\mbox{\includegraphics[width=3.0in]{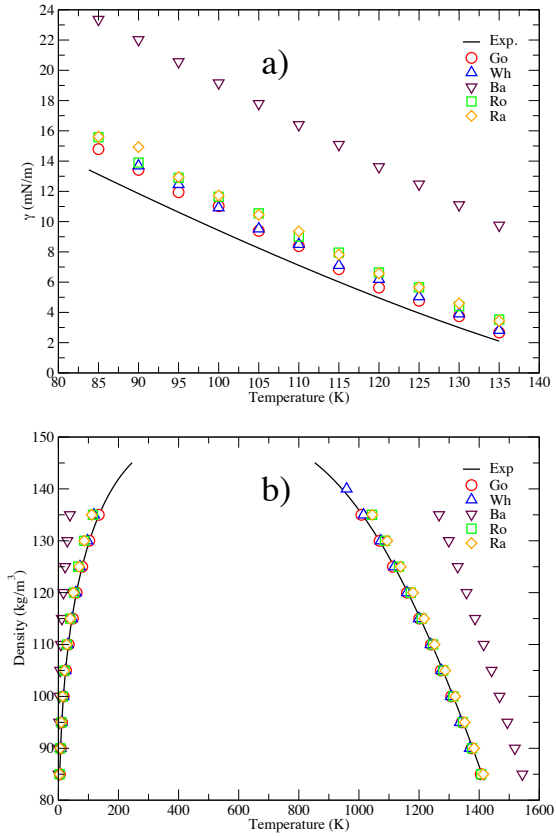}}
\caption{a) Liquid-vapor phase diagram and b) surface tension as a function of temperature 
for different Lennard-Jones parameters. The experimental results in both graphs are shown  
as a solid line \cite{experiment}.}
\label{fig1}
\end{figure}

%\ioptwocol
\noindent Reparameterization is an empirical (fast and easy) method to obtain better properties 
of modeled compounds by scaling charges, Lennard-Jones (LJ) energy and distance parameters, and 
bond distance to reach the dielectric constant, surface tension, density or micelle radius 
\cite{reparam,tensoactives}, and self-diffusion constant \cite{propanol}, respectively. There 
are many argon force fields that reproduce the liquid-vapor phase diagram, but the surface 
tension is not obtained with the same force field \cite{tildesley,white,rahman,barker,rowley}. 
In Figure \ref{fig1} we can appreciate different force fields; the force field named Go, 
developed by Goujon et al. \cite{tildesley}, was obtained by considering an extra quadrupolar 
term to the LJ potential, this force field has been improved to obtain both the liquid-vapor 
phase diagram and surface tension, adding to last property the effect of three-body 
interactions. The force field developed by White (Wh) is based on a renormalization group 
theory \cite{white}, the phase diagram fits excellently with the experiment except for surface 
tension. Similar behaviors to the one mentioned above are the force fields developed by Barker 
et al. \cite{barker}, Rowley et al. \cite{rowley}, and Rahman \cite{rahman}, labeled as Ba, Ro, 
and Ra, respectively. 

\begin{figure}
\centering\mbox{\includegraphics[width=3.0in]{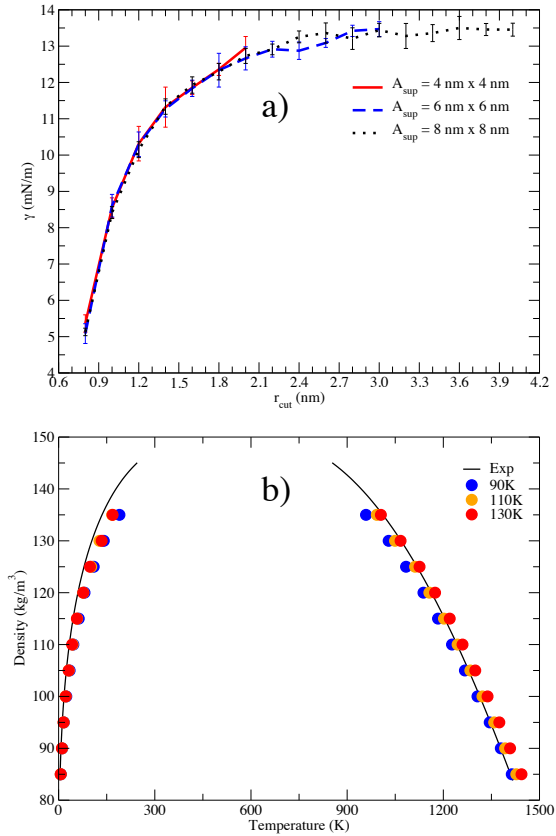}}
\caption{a) Surface tension in different interfacial area by varying cut radium, 
b) reparameterization of the argon parameters for different temperatures, 90K, 110K 
and, 130K. The surface tension experimental values are shown as the solid line
 \cite{experiment}.}
\label{fig2}
\end{figure}

To avoid any finite size effects to calculate the surface tension, we analize different 
interfacial areas and cut radius ($r_{\mbox{cut}}$) as illustrated in figure \ref{fig2} 
a), where the surface tension values for an area of 4 nm x 4 nm are not stable, these values 
increase as the $r_{\mbox{cut}}$ increases, for the area of 6nm x 6 nm they are stable from 
$r_{\mbox{cut}}$ = 2.8 nm, and for 8 nm x 8 nm stability is reached from 2.6 nm. Then, 
molecular simulation was performed at $r_{\mbox{cut}}$ = 2.8 nm with an interfacial area of 
6 nm $\times$ 6 nm and 30 nm in Z direction containing $N = 6750$ argon atoms. This analysis 
and the scaling process improved the value of the surface tension, and avoiding the addition 
of three-body interaction to the surface tension to reach its experimental value, as applied 
by Goujon et al. \cite{tildesley}.

The molecular dynamics simulation was carried out using GROMACS software \cite{gromacs} with 
the Velocity-Verlet algorithm, an NVT ensemble was used to obtain the surface tension and 
density profile. A V-rescale thermostat was used ($\tau$ = 0.5 ps) to keep the temperature 
constant. Furthermore, a time step of $\Delta t$ = 0.002 ps and periodic boundary conditions were 
established. Zero charge and molecular weight of 39.948 was used. All simulations were 
run 5 ns of production after 5ns of equilibration. The surface tension was calculated from 
the mechanical definition,

\begin{equation}
\gamma=\frac{1}{2} L_z \left[ P_{zz}-\frac{1}{2}(P_{xx}+P_{yy}) \right].
\end{equation} 

The temperature to perform reparameterizations was kept constant, after obtaining surface 
tension and density of the liquid; the temperature was varied to obtain the complete 
liquid-vapor phase diagram and the temperature dependence of the surface tension. To perform 
the reparameterization, a temperature value of 110K was chosen to ensure that the simulations 
data fit as best as possible over the entire temperature range. We reached that conclusion by 
taking the force field from Goujon et al. \cite{tildesley}, then reparameterizing the force 
field at different temperatures, figure \ref{fig2} shows the reparameterized data at 90 K, 
110 K, and 130 K. The data obtained at a temperature of 110 K are closer to the experimental 
liquid density than other temperatures, so the other force fields were reparameterized at that 
temperature. 

\begin{figure}
\centering\mbox{\includegraphics[width=3.0in]{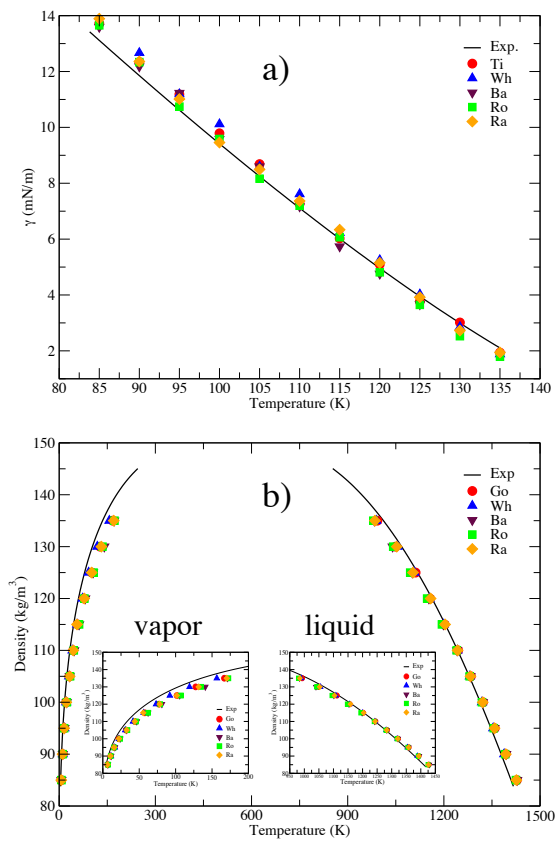}}
\caption{Reparametrized a) liquid-vapor phase diagram and b) surface tension as a 
function of temperature for different Lennard-Jones parameters. The experimental 
results in both graphs are shown as solid line \cite{experiment}.}
\label{fig3}
\end{figure}

The reparameterization procedure was applied to obtain the surface tension and the 
liquid-vapor phase diagram at a fixed temperature. The temperature dependence of these 
properties is shown in Figure \ref{fig3}, it is observed that all force fields are close to 
the experimental surface tension as a function of temperature, the Wh force field is closer 
than the other force fields \cite{white} for both properties, see insets in Figure \ref{fig3}, 
this is independent of how the original parameters were obtained and depends on the ability 
to adjust them to obtain the experimental properties. The phase diagram has a similar behavior 
for almost all force fields; the liquid density branches are well described, although the 
vapor line is not as good as the Wh force field, the surface tension and phase diagram are 
well reproduced by this model after reparameterization. 

\begin{figure}
\centering\mbox{\includegraphics[width=3.0in]{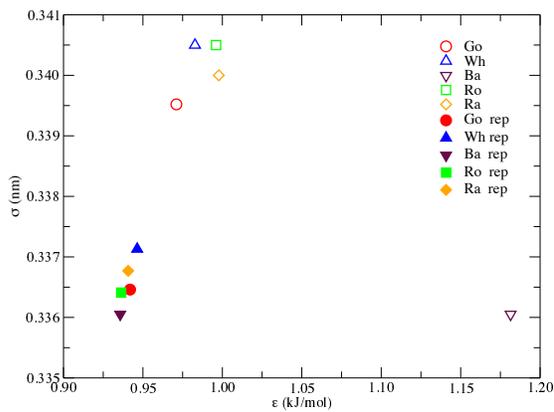}}
\caption{$\epsilon$-$\sigma$ LJ parameters phase diagram.}
\label{fig4}
\end{figure}

The Lennard-Jones parameters fron literature and the reparameterized ones (obtained in this 
work) are summarized in Table \ref{forcefield1}. The new or reparametrized parameters are 
close to each others, but far from the original ones that only reproduce the liquid-vapor 
phase diagram, Figure \ref{fig4}, both force fields represent separate regions in the 
$\epsilon$-$\sigma$ phase space, there is an original force field (Ba) \cite{barker} which 
is not part of any region, this force field did not reproduce both properties, this phase 
space indicates that, if we want to reproduce the liquid-vapor phase diagram, we have to 
chose a value of $\epsilon$ from 0.97107 kJ/mol to 0.99774 kJ/mol and an $\sigma$ value 
between 0.33605 nm to 0.34050 nm, if we also want to reproduce the surface tension then we 
have to chose a value of $\epsilon$ from 0.94191 kJ/mol to 0.94637 kJ/mol and a velue of 
$\sigma$ from 0.33605 nm to 0.33713 nm of $\sigma$.

\begin{table}
\begin{center}
%\scriptsize{
\caption{Left table) Literature and right table) reparameterized $\epsilon$ and 
$\sigma$ Lennard-Jones parameters for argon compound in kJ/mol and nm, respectively.}
\label{forcefield1}
\begin{tabular}{lrr}
\hline
\hline
Ref.              &  $\epsilon$  & $\sigma$  \\
\hline
\hline
 Ti=\cite{tildesley}  &  0.97107   & 0.33952    \\
 Wh=\cite{white}      &  0.98275   & 0.34050    \\
 Ba=\cite{barker}     &  1.18144   & 0.33605    \\
 Ro=\cite{rowley}     &  0.99607   & 0.34050    \\
 Ra=\cite{rahman}     &  0.99774   & 0.34000    \\
\hline
\hline
\end{tabular}
\begin{tabular}{lrr}
\hline
\hline
Ref.              &  $\epsilon$  & $\sigma$  \\
\hline
\hline
 Ti=\cite{tildesley}  &  0.94191   & 0.33646    \\
 Wh=\cite{white}      &  0.94639   & 0.33713    \\
 Ba=\cite{barker}     &  0.93570   & 0.33605    \\
 Ro=\cite{rowley}     &  0.93631   & 0.33641    \\
 Ra=\cite{rahman}     &  0.94071   & 0.33677    \\
\hline
\hline
\end{tabular}
%}
\end{center}
\end{table}

The empirical process used in this work could be improved by obtaining the LJ parameters by 
a renormalization group theory, as in the Wh model, which was obtained \cite{white,white_org} 
by applying it not only to liquid-vapor phase diagram prediction, but also to the surface 
tension. The original methodology considers the contribution of the repulsive potential only 
for high temperatures and for low temperatures the part of the attractive potential, 
considering a renormalization procedure to the Helmholtz free energy, this expression and a 
correct choice of parameteres ($\epsilon$, $\sigma$) provides a complete prediction of the 
liquid-vapor phase diagram. 

\ack 

The authors acknowledge the computer facilities from Laboratorio Nacional de 
Superc\'omputo del Suroeste de M\'exico (LNS) project 201801014N1R and 201901004N. 
This work was partially supported by grant CB-2016-01 286238 CONACyT, M\'exico.

\end{document}